\documentclass[11pt,moriond,epsfig]{article}
\usepackage{moriond,epsfig}

\def\be{\begin{equation}}
\def\ee{\end{equation}}
\def\bea{\begin{eqnarray}}
\def\eea{\end{eqnarray}}

\begin{document}
\vspace*{4cm}
\title{CLUSTERS OF GALAXIES AMONG ROSAT BLANK FIELD SOURCES}

\author{I. CAGNONI$^{1,2}$, M. ELVIS$^2$, D.-W. KIM$^2$, P. MAZZOTTA$^2$, J.-S. HUANG$^2$  \& A. CELOTTI$^1$}

\address{$^1$ SISSA-ISAS, Via Beirut 2-4, 34014 Trieste, Italy\\
$^2$ SAO, 60 Garden St., 02138 Cambridge, MA, USA}

\maketitle\abstracts{We present here an efficient method for 
selecting high luminosity and massive high redshift clusters
of galaxies,  crucially important tools in cosmology.
By selecting bright and extremely X-ray loud  (high $F_X/F_V$) sources,
we were able to identify  2 high redshift ($z=0.45$ and $z=0.89$)
clusters so far and we have evidence for at least one more candidate 
at $0.4<z<1.1$ out of a total of 16 selected sources.}

\section{Importance of high z clusters}

High luminosity  and massive high redshift clusters of galaxies
are crucially important tools in cosmology.
Their distribution and evolution is  fully determined by the spectrum
of primordial perturbations and cosmological parameters $\Omega_0$ and 
$\Lambda$ (e.g. Press \& Schechter 1974). 
In particular, the models of low $\Omega$ universe (with or without cosmological 
constant) (e.g. Henry 2000, Borgani \& Guzzo 2001) predict a higher density 
of massive clusters  at high redshifts than  the high $\Omega$ models.\\
A sample of high redshift clusters is essential in
determining the evolution of the clusters X-ray luminosity function 
(e.g. Rosati et al., 1998),  and also 
would allow us to determine if there is  evolution in the 
luminosity-temperature ($L_X-T$) relation. The evolution of the $L_X-T$ is
very important since it is related to physical mechanisms of cooling and heating in the central
 cluster region (e.g. Tozzi \& Norman, 2001).
Only a few bright high redshift clusters have been found so far (e.g. in the EMSS, Gioia \& Luppino 1994; in the RDCS Rosati et al. 1999, Della Ceca et al. 2000; in the WARPS Ebeling et al. 1998; 2000) and only 9 of them at $z>0.5$ have a measure of their temperature
(e.g. Della Ceca et al. 2001, Cagnoni et al. 2001, Stanford et al. 2001).
Since the statistics is very scanty, the search for high redshift and high luminosity clusters is the only means to improve such studies.
Two high redshift ($z \geq 0.45$) clusters of galaxies have been already
 found among the 16 {\em ROSAT}  blank field sources and other can be present in the remaining 11 unidentified sources.
We will describe the selection method in Section 2, and discuss the possibilities regarding the nature of the blanks and the presence of high redshift clusters of galaxies among them in Section 3.

\section{Blank field {\em ROSAT} sources}

We call  `blank field sources' (blanks) all the bright X-ray sources 
($F_X > 10^{-13}$ erg cm$^{-2}$ s$^{-1}$) with no optical counterpart 
on the Palomar Sky survey (to O=21.5) within
their $39^{\prime \prime}$ (99\%) radius error circle.\\
To define our sample of blank field sources
we searched the {\it ROSAT} pointed archive (WGACAT, White Giommi \& Angelini 
1984) with the following selection criteria:
\begin{enumerate}
\item bright X-ray sources with $f_X >10^{-13}$ erg cm$^{-2}$ s$^{-1}$;
\item good detection (SNR $>10$);
\item high Galactic latitude ($|b| > 20^{\circ}$);
\item location in the `inner circle' of the PSPC ($r<18^{\prime}$) to
have smaller positional error circles.
\end{enumerate}
We then used the APM digitization of the Palomar and UK Schmidt
sky surveys (McMahon \& Irwin, in preparation), via the WWW, to
find sources with no optical counterpart within the 99\% X-ray
position circle ($r\sim 39^{\prime\prime}$).
Due to a problem with rev~0 images used for WGACAT
 (Giommi, Angelini and Cagnoni, private communication 1999) 
$\sim 35$\% of the blanks had wrong coordinates and were not actually blank.
We quote here the correct positions based on rev~2 version 
(from the Galpipe catalog) of ROSAT data.
The total sample consists of 16 sources.

\section{Clusters of galaxies among ROSAT blanks}

Blanks are extreme X-ray loud objects: 
no known extragalactic population has an appropriate $F_X/F_V$ (see Maccacaro
et al. 1988 nomograph).
Normal quasars and AGN have 0.1$< F_X/F_V <$10; BL~Lac objects
can reach $F_X/F_V$=40; normal galaxies have 10$^{-4}<
F_X/F_V <$10$^{-2}$; while a high luminosity cluster of galaxies
($L_X=10^{45}$ erg~s$^{-1}$) would have a first ranked elliptical
with $L_{gal}=10L^{\ast} (=10^{11.5}L_{\odot}$) giving $F_X/F_V =
8$. None of these reach the $F_X/F_V > 60$ of the blanks we have found. \\
The still open possibilities regarding the nature of blanks are:\\
1) - Quasar~2s: high luminosity, high redshift heavily obscured quasars, 
the bright analogs of the well known Seyfert 2s;\\
2) - low mass Seyfert~2: that is AGNs powered by a low mass obscured black hole
(i.e. obscured Narrow Line Seyfert 1);\\
3) - high redshift clusters of galaxies;\\
4) - failed clusters: in which a large overdensity of matter has collapsed but 
has not formed galaxies;\\
5) - AGNs with no big blue bump.\\
The faintness of their optical counterparts and the flatness of their PSPC
 spectra can be, in cases (1) and (2), due to the effect of the intrinsic 
absorption.
In the case of high redshift cluters, the flattness of the PSPC spectra could be due to a high temperature thermal spectrum and the faintness in the optical
to the shift of the 4000 \AA \/ break at longer wavelenghts 
than the Palomar O band.

In both cases we would expect these blanks to show extremely red colors  (e.g. Webster et al. 1995; Songaila et al. 1994); we thus obtained optical and IR imaging for all the 16 sources 
in R and K to R=22 (this corresponds
to B$\sim$23.5 for a galaxy) and K=20.
Indeed 6 sources show red  colors and 3 of them clearly show an excess of IR sources in the field of view. 
Two of these red sources were identified as high redshift clusters of galaxies. We will detail about each of them in the following.

\subsection{1WGA~J1226.9$+$333}

We obtained a {\em Chandra} 10~ks observation for 1WGA~J1226.9$+$333 (Cagnoni et al. 2001) and the source clearly looks extended on an arcminute scale with a circularly symmetrical and smooth profile (Fig.~1).
Combining this observation with the constraints from optical and IR images we derived a redshift of $z=0.85 \pm 0.15$ and a $T\sim 10$ keV.
These results perfectly agree with a spectroscopic measure of the redshift 
$z=0.888$ (Ebeling et al. 2001) and with a Sunyaev-Zel'dovich
 measurement of the temperature by Joy et al. (2001) ($kT =10.0^{+2.0}_{-1.5}$~keV).
For z=0.89 the unabsorbed bolometric and 0.5-2.0 keV band luminosities are 
 $L=(2.2 \pm 0.2) \times 10^{45}$ erg~s$^{-1}$ and  $L_{X(0.5-2.0)}=(4.4 \pm 0.5) 
\times 10^{44}$ erg~s$^{-1}$ respectively (assuming H$_0$=75~km~s$^{-1}$~Mpc$^{-1}$ 
and $q_0 = 0.5$).

This makes 1WGA~J1226.9$+$3332 one of the most luminous clusters at high redshift and it is a statistically important point for cosmological studies.

\begin{figure}
\begin{center}
\psfig{figure=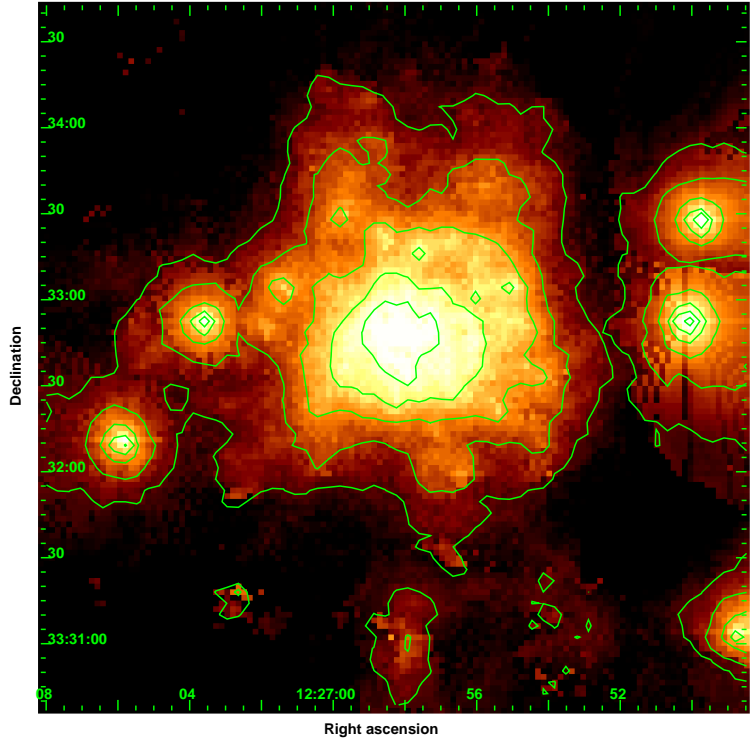,height=4.0in}
\caption{{\em Chandra} adaptive smoothed and background subtracted image of the cluster  1WGA~J1226.9$+$333 extracted in the 0.5-5.0 keV energy range.}
\end{center}
\end{figure}

\subsection{1WGA~J0221.1$+$1958}

1WGA~J0221.1$+$1958 was identified in the SHARC survey (Romer et al. 2000)
as a $z=0.45$ cluster of galaxies, and assuming a $T\sim 6$ keV, they derived
a {\em ROSAT} X-ray luminosity of 
$L_{X(0.5-2.0)}=2.87 \times 10^{44}$  erg~s$^{-1}$ (H$_0$=50~km~s$^{-1}$~Mpc$^{-1}$ 
and $q_0 = 0.5$).

\subsection{A candidate high z cluster}

There is a nother object which shows all the properties
 of the $z=0.89$  high luminosity cluster of galaxies 
1WGA~J1226.9$+$3332.
In particular both for this object and for 1WGA~J1226.9$+$3332
 there is an IR bright (K$\sim 15.8$) red 
(R-K$\sim 5.2$) galaxy in the X-ray error circle and  a strong 
($> 7\sigma$ at K=18) excess of extended and red (R-K$\geq 4.5$) sources.
If this galaxy were a first ranked cluster ellipticals with 
 $M_K =-26.7 \pm 0.5$ (assuming negligible K-correction,  $q_0 =0.5$ and , 
H$_0$=50~km~s$^{-1}$~Mpc$^{-1}$ as in  Collins \& Mann, 1998), 
 we would have a limit on the redshift: $z=0.85^{+0.40}_{-0.25}$
 for the candidate cluster
 ($z=0.68^{+0.30}_{-0.19}$ from the K=15.5 galaxy in 1WGA~J1226.9$+$3332 field of view).
This galaxy is extremely red ($R-K \sim 5.1 \pm 0.5$ 
for the candidate cluster
and $5.1 \pm 0.3$ 
for 1WGA~J1226.9$+$3332), as expected for and high z clusters of galaxies.
Only an unevolving elliptical galaxy at $0.7<$z$<1.1$
 or a Sbc galaxy at z$>1.2$ can have such red color
 (Coleman, Woo \& Weedman, 1980).
($0.75 < z < 1.0$ are the limits for the high z cluster 1WGA~J1226.9$+$3332).

This observational evidence strongly support the identification of this object with a high redshift ($0.5 \leq z \leq 1.0$) cluster of galaxies.

\vspace{0.2in}

Thanks to the large area sampled and with two high redshift ($z=0.45$ and $z=0.89$) clusters of galaxies and one more candidate out of the 16 selected sources, the selection of bright and extremely X-ray loud sources
 ({\em ROSAT} blank field sources) proved to be the most efficient method for finding highly luminous and high redshift clusters of galaxies.

\section*{Acknowledgments}

I.C. would like to thank the organizers of the meeting
 for the financial support.
This work was supported by NASA grant GO 0-1086X
 and by the Italian MURST (IC and AC).
P.M. acknowledges an ESA fellowship.


\section*{References}
\begin{thebibliography}{99}


\bibitem{} Borgani \& Guzzo, 2001, Nature, 409, 68016, 39

\bibitem{} Cagnoni, I., Elvis, M., Kim, D.-W., Mazzotta, P., Huang, 
J.-S. \& Celotti, A., 2001, ApJ in press

\bibitem{}Collins, C.A. \& Mann, R. G., 1998, MNRAS, 297, 128

\bibitem{}Coleman, G.D., Wu, C.-C. \& Weedman, D.W., 1980, ApJS, 43, 393

\bibitem{}Della Ceca, R., Scaramella, R., Gioia, I.M., Rosati, P., Fiore, F. \& Squires, G., 2000, A\&A, 353, 498

\bibitem{} Gioia, I. M. \& Luppino, G. A., 1994, ApJS, 94, 538

\bibitem{} Ebeling, H. et al. 1998, MNRAS, 301, 881

\bibitem{} Ebeling, H. et al. 2000, MNRAS, 534, 133

\bibitem{} Ebeling, H. et al. 2001, ApJ, 548, 23

\bibitem{}Henry, J.\ P.\ 2000, ApJ, 534, 565

\bibitem{} Joy, M. et al., 2001, ApJL submitted (astro-ph/0012052)

\bibitem{}Maccacaro, T.,   Gioia, I.M., Wolter, A., Zamorani, G. \&
 Stocke, J.T., 1988, ApJ, 326, 680

\bibitem{} Press, W.\ H.\ \& 
Schechter, P.\ 1974, ApJ, 187, 425 

\bibitem{} Romer et al, 2000, ApJS, 126, 209

\bibitem{} Rosati, P., della Ceca, R., Norman, C.\ \& Giacconi, R.\ 1998, ApJL, 492, L21 

\bibitem{} Rosati, P. et  al. 1999, AJ, 118, 26

\bibitem{}  Songaila, A. et al. 1994, ApJS, 94, 461

\bibitem{}Stanford, S. A. et al,  2001, ApJ in press (astro-ph/0012250)

\bibitem{} Tozzi \& Norman, 2001, ApJ, 546, 63

\bibitem{} Webster, R. L., 1995, Nature, 375, 469

\bibitem{}White, N.E., Giommi, P. \& Angelini, L., 1994, IAUC 6100

\end{thebibliography}
\end{document}